\documentclass[%
 reprint,
superscriptaddress,
 amsmath,amssymb,
 aps,
prl,
]{revtex4-2}

\usepackage{graphicx}
\usepackage{dcolumn}
\usepackage{bm}
\usepackage{siunitx}

\begin{document}

\preprint{APS/123-QED}

\title{Perpendicular subcritical shock structure in a collisional plasma experiment}

\author{D. R. Russell}
\email[]{daniel.russell13@imperial.ac.uk}
\affiliation{Blackett Laboratory, Imperial College London, London SW7 2AZ, United Kingdom}

\author{G. C. Burdiak}
\affiliation{First Light Fusion Ltd, Yarnton, Kidlington OX5 1QU, United Kingdom}

\author{J. J. Carroll-Nellenback}
\affiliation{Department of Physics and Astronomy, University of Rochester, Rochester, NY 14627, USA}

\author{J. W. D. Halliday}
\affiliation{Blackett Laboratory, Imperial College London, London SW7 2AZ, United Kingdom}

\author{J. D. Hare}
\affiliation{Plasma Science and Fusion Center, Massachusetts Institute of Technology, Cambridge MA 02139,
USA}

\author{S. Merlini}
\affiliation{Blackett Laboratory, Imperial College London, London SW7 2AZ, United Kingdom}

\author{L. G. Suttle}
\affiliation{Blackett Laboratory, Imperial College London, London SW7 2AZ, United Kingdom}

\author{V. Valenzuela-Villaseca}
\affiliation{Blackett Laboratory, Imperial College London, London SW7 2AZ, United Kingdom}

\author{S. J. Eardley}
\affiliation{Blackett Laboratory, Imperial College London, London SW7 2AZ, United Kingdom}

\author{J. A. Fullalove}
\affiliation{Blackett Laboratory, Imperial College London, London SW7 2AZ, United Kingdom}

\author{G. C. Rowland}
\affiliation{Blackett Laboratory, Imperial College London, London SW7 2AZ, United Kingdom}

\author{R. A. Smith}
\affiliation{Blackett Laboratory, Imperial College London, London SW7 2AZ, United Kingdom}

\author{A. Frank}
\affiliation{Department of Physics and Astronomy, University of Rochester, Rochester, NY 14627, USA}

\author{P. Hartigan}
\affiliation{Department of Physics and Astronomy, Rice University, Houston, TX 77005-1892, USA}

\author{A. L. Velikovich}
\affiliation{Plasma Physics Division, Naval Research Laboratory, Washington, DC 20375, USA}

\author{S. V. Lebedev}
\affiliation{Blackett Laboratory, Imperial College London, London SW7 2AZ, United Kingdom}

\date{\today}

\begin{abstract}
We present a study of perpendicular subcritical shocks in a collisional laboratory plasma. Shocks are produced by placing obstacles into the super-magnetosonic outflow from an inverse wire array z-pinch. We demonstrate the existence of subcritical shocks in this regime and find that secondary shocks form in the downstream. Detailed measurements of the subcritical shock structure confirm the absence of a hydrodynamic jump. We calculate the classical (Spitzer) resistive diffusion length and show that it is approximately equal to the shock width. We measure little heating across the shock ($< 10 \%$ of the ion kinetic energy) which is consistent with an absence of viscous dissipation.
\end{abstract}

\maketitle

Shock waves are ubiquitous in astrophysical \cite{Treumann2009}, space \cite{Burgess2015} and laboratory plasmas and often include an embedded, dynamically significant magnetic field. The theoretical understanding of magneto-hydrodynamic (MHD) shock waves was first established in the 1950s \cite{Hoffmann1950,Germain1960,Polovin1961,Anderson1963}. In particular, it was discovered that resistive Ohmic heating, a dissipation mechanism specific to MHD, can shape shock structures. Since resistivity does not directly dissipate the plasma kinetic energy, there is a critical value of the upstream magnetosonic Mach number, $M_{C}$, indicating the maximum strength of MHD shocks shaped by Ohmic heating alone. This is defined by setting the downstream sonic Mach number $M_{S, d} = 1$ \cite{Marshal1955,Coroniti1970}. The downstream flow is supersonic for subcritical shocks, and hydrodynamic parameters are continuous across the shock.

Since then, studies of subcritical shocks have focused on the collisionless regime \cite{Mellott1984a,Schaeffer2015}, motivated by the ubiquity of collisionless shocks in space and astrophysical plasmas. In these cases, the physical processes which generate entropy at the shock are more complicated than the collisional transport models \cite{Braginskii1965,Epperlein1998}. This leads to difficulty in defining the critical Mach number (due to the lack of downstream thermodynamic equilibrium) and determining the plasma resistivity. Recent progress in magneto-inertial fusion \cite{Wurden2015,Slutz2010,Gomez2020,Perkins2017}, has stimulated renewed interest in MHD shocks propagating through dense, collision-dominated plasmas. In particular, recent experiments studying the implosion of magnetised inertial confinement fusion capsules have shown an increased yield and anisotropic shock structure \cite{Moody2021,Ho2021}. The structures of MHD shocks in this regime have been investigated theoretically \cite{Kennel1988,Liber1994} but never measured experimentally.

This letter reports the first experimental study of subcritical shock structure in a highly collisional plasma (m.f.p. $<<$ shock width). A supersonic ($M_{S} \sim 2.5$), super-Alfv\'{e}nic ($M_{A} \sim 3$) plasma flow was produced by the current driven ablation of an inverse wire array z-pinch \cite{Harvey-Thompson2009}, and shocks were studied by placing stationary obstacles into this flow \cite{Burdiak2017}. The orientation of the obstacles produced perpendicular shocks, in which the advected magnetic field was perpendicular to the shock normal. Subcritical shocks were observed in which the downstream flow is shown to be supersonic. These measurements of a subcritical shock in a collisional plasma are the first of their kind and confirm the absence of a hydrodynamic jump as predicted by theory (e.g. \cite{Coroniti1970,Liber1994}). Furthermore, detailed laser probing measurements allow calculation of the resistive diffusion length, $L_{\eta}$ (using classical Spitzer resistivity) which we show to be approximately equal to the shock width. The shock structure can therefore be described by classical resistive MHD, without the inclusion of anomalous resistivity. We observed little plasma heating at the shock ($< 10 \%$ of the ion kinetic energy) consistent with a balance between adiabatic and Ohmic heating and radiative cooling.

\begin{figure}
	\includegraphics[scale=1]{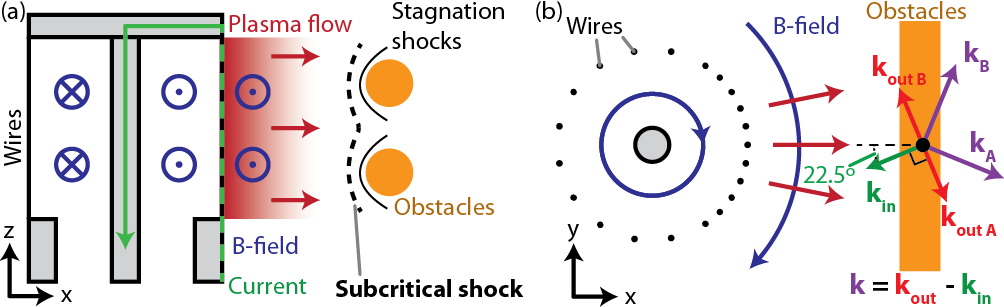} 
	\centering
	\caption{Cross-sectional diagram of the wire array and obstacles. (a) Side-on view. Cylindrical obstacles extend into the page and shocks are indicated where they form in the experiment. (b) End-on view showing the azimuthal configuration of wires in the array and the vector diagram for the Thomson scattering diagnostic.}
	\label{fig:Setup}
\end{figure}

The experimental setup is shown in Fig. \ref{fig:Setup}. The inverse wire array z-pinch was driven by the MAGPIE pulsed power generator (1.4 MA peak current, 240 ns rise time) \cite{Mitchell1996}. A cylindrical arrangement (21 mm high and 20 mm in diameter) of 21 aluminium wires (each with a 40 \si{\micro\metre} diameter) surrounded a central cathode. This experimental geometry provided a $\bf{J} \times \bf{B}$ force which acted radially outwards, accelerating the plasma ablated by the wires for the duration of the drive current \cite{Lebedev2014}. Since some of the drive current passed through the ablated plasma surrounding each wire, a fraction of the magnetic field was advected by the flow \cite{Chittenden2004}. The 11 wires closest to the obstacles had an angular separation of $11.25^{\circ}$ while the remaining wires had an angular separation of $22.5^{\circ}$, see Fig. \ref{fig:Setup} (b). The smaller wire separation reduced azimuthal density variation by reducing the divergence of the flow and allowing outflows from adjacent wires to merge \cite{Burdiak2017}.

Two 4 mm diameter cylindrical brass obstacles were placed 10 mm from the ablating wires and were oriented with their axes parallel to the advected magnetic field (Fig. \ref{fig:Setup}). The obstacles had a centre-to-centre separation of 9 mm in the vertical, $z$, direction and were 40 mm in length. The obstacles produced shocks in the plasma flow which were extended in the $y$ direction and approximately stationary in the laboratory frame for $\sim 200$ ns (many hydrodynamic crossing times).

\begin{figure}
	\includegraphics[scale=1]{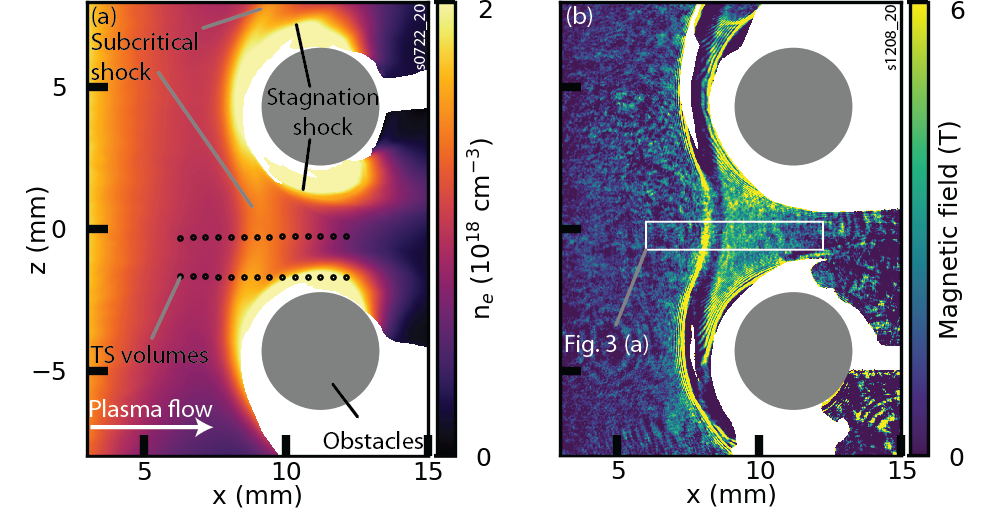}
	\centering
	\caption{Electron density and magnetic field measurements. (a) Electron density recorded $426$ ns after current start. The edge of the wire array is at $x = 0$ mm. (b) Magnetic field $397$ ns after current start. The region sampled by the lineout in Fig. \ref{fig:TS} is shown.}
	\label{fig:int_far}
\end{figure}

Electron density was measured using a Mach-Zehnder laser interferometer ($532$ nm, $0.4$ ns FWHM). The analysis method is described in Refs. \cite{Swadling2014, Hare2019}. An electron density map in the $x$-$z$ plane is presented in Fig. \ref{fig:int_far} (a). The line averaged electron density was calculated by $n_{e} \approx \int n_{e} dy /L$ where the choice $L = 40$ mm is consistent with interferometry measurements in the $x$-$y$ plane.

The observed shock structure comprises two distinct types of shock. A subcritical shock spans the region upstream of both obstacles. This shock is smooth and continuous between the obstacles. Fast-frame optical self-emission imaging has shown that this shock forms as a single shock $\sim 300$ ns after current start \cite{Lebedev2014,Burdiak2017}. In addition to the subcritical shock, two secondary shocks, referred to here as stagnation shocks, form closer to the obstacles, downstream of the subcritical shock. These do not expand upstream to reach the position of the subcritical shock. The stagnation shocks cause a more abrupt density increase than the subcritical shock.

The magnetic field distribution in the $x$-$z$ plane was measured using a Faraday rotation imaging diagnostic in combination with inline interferometry ($1053$ nm, $1$ ns FWHM). The diagnostic is described in Ref. \cite{Swadling2014} and the analysis in Ref. \cite{Russell2021}. The measurement is sensitive to the line averaged, electron density weighted $B_{y}$ component of the magnetic field. This is the dominant component in the shock region, however, we expect some curvature of the magnetic field due to the cylindrical structure of the wire array, so Faraday rotation imaging provides a lower bound for $B_{y}$ at $y=0$ mm. Fig. \ref{fig:int_far} (b) shows the measured magnetic field. In the region between the obstacles, the magnetic field increases from $1.5 - 2$ T upstream of the subcritical shock to $3 - 4$ T in the downstream, showing that magnetic field is compressed across the shock. Redistribution of the laser intensity caused by refraction at density gradients (shadowgraphy) leads to an intensity modulation at the subcritical shock ramp so magnetic field cannot be inferred inside the shock.

Optical Thomson Scattering (TS) of the ion acoustic feature provided localised measurements of plasma velocity and temperatures \cite{Swadling2014,Suttle2021}. A focused laser beam ($532$ nm, $8$ ns FWHM, $200$ \si{\micro\metre} beam waist) passed between the obstacles at $22.5 ^{\circ}$ to the $x$-axis in the $x$-$y$ plane (\textbf{k\textsubscript{in}} in Fig. \ref{fig:Setup} (b)). The scattered light was imaged onto two separate arrays of 14 optical fibres at $\pm 90 ^{\circ}$ to the laser beam (\textbf{k\textsubscript{out A}} and \textbf{k\textsubscript{out B}}), which recorded the scattered spectra from the same 14 localised plasma volumes (shown in Fig. \ref{fig:int_far} (a)). The Doppler shifted spectra are sensitive to the velocity components along \textbf{k\textsubscript{A}} and \textbf{k\textsubscript{B}}. Temperatures and velocities were inferred by fitting theoretical spectra to the experimental data \cite{Froula2011,Hare2017}.

The flow velocity directly upstream of the subcritical shock was $45$ \si{\kilo\metre\per\second} and the temperature was $T_{e} = T_{i} = 11 \pm 3$ eV (self consistent values of $T_{e}$ and $\bar{Z}$ were inferred from the experimentally measured $\bar{Z}T_{e}$ using an nLTE atomic model \cite{Niasse2011}). The fact that we measure $T_{e} = T_{i}$ is unsurprising since the thermal equilibration time $\tau_{i/e} \sim 5$ ns.

Combined analysis of interferometry, Faraday rotation and TS data allow characteristic dimensionless parameters of the upstream plasma to be calculated, see Table \ref{flow_parameters}. The thermal and magnetic pressures differ by less than a factor 2, while the ram pressure is substantially larger. The Mach numbers all exceed unity so the flow will form a shock when colliding with stationary obstacles. The critical Mach number (which depends on $\beta_{th}$ and the shock angle \cite{Edmiston1984}) for these upstream parameters is $M_{C} \sim 1.4$ (assuming $\gamma = 5/3$). However, we note that only a small increase in the magnetic field, to $2.2$ T, would be required for the flow to be subcritical with respect to a stationary shock ( giving $M_{MS} = 1.7$ and $\beta_{th} = 1$). Since the magnetic field inferred from the Faraday rotation data gives a lower bound for $B_{y}(y=0$ mm$)$, we conclude that a subcritical shock may form. The large Reynolds number means viscous dissipation will occur on scales much smaller than the system size. However, the modest value of $Re_{M}$ shows that while magnetic field is expected to be advected in the upstream, magnetic diffusion will become important on the spatial scale of the shocks.

\begin{table}
	\begin{tabular}{ p{5cm} p{1.3cm} p{1.8cm} }
	\hline
	\hline
	Dimensionless parameter & & Value \\
	\hline
	Thermal beta & $\beta_{th}$ & $1.7$ \\
	Dynamic beta & $\beta_{ram}$ & $18$ \\
	Sonic Mach number & $M_{S}$ & $2.5$ \\
	Alfv\'{e}nic Mach number & $M_{A}$ & $3$ \\
    Magnetosonic Mach number & $M_{MS}$ & $1.9$ \\
	Reynolds number & $Re$ & $4 \times 10^{4}$ \\
	Magnetic Reynolds number & $Re_{M}$ & $10 \to 1$ \\
	\hline
	\hline
	\end{tabular}
	\centering
	\caption{Characteristic plasma parameters upstream of the subcritical shock $\sim 400$ ns after current start. $\beta_{th} = P_{th}/P_{mag}$ and $\beta_{ram} = P_{ram}/P_{mag}$. To evaluate the magnetic Reynolds number, a scale length of $10$ mm, the distance between the wire array and the obstacles, gives $Re_{M} \sim 10$ while a distance of $\sim 0.8$ mm, the subcritical shock width, gives $Re_{M} \sim 1$.}
	\label{flow_parameters}
\end{table}

\begin{figure}
	\includegraphics[scale=1]{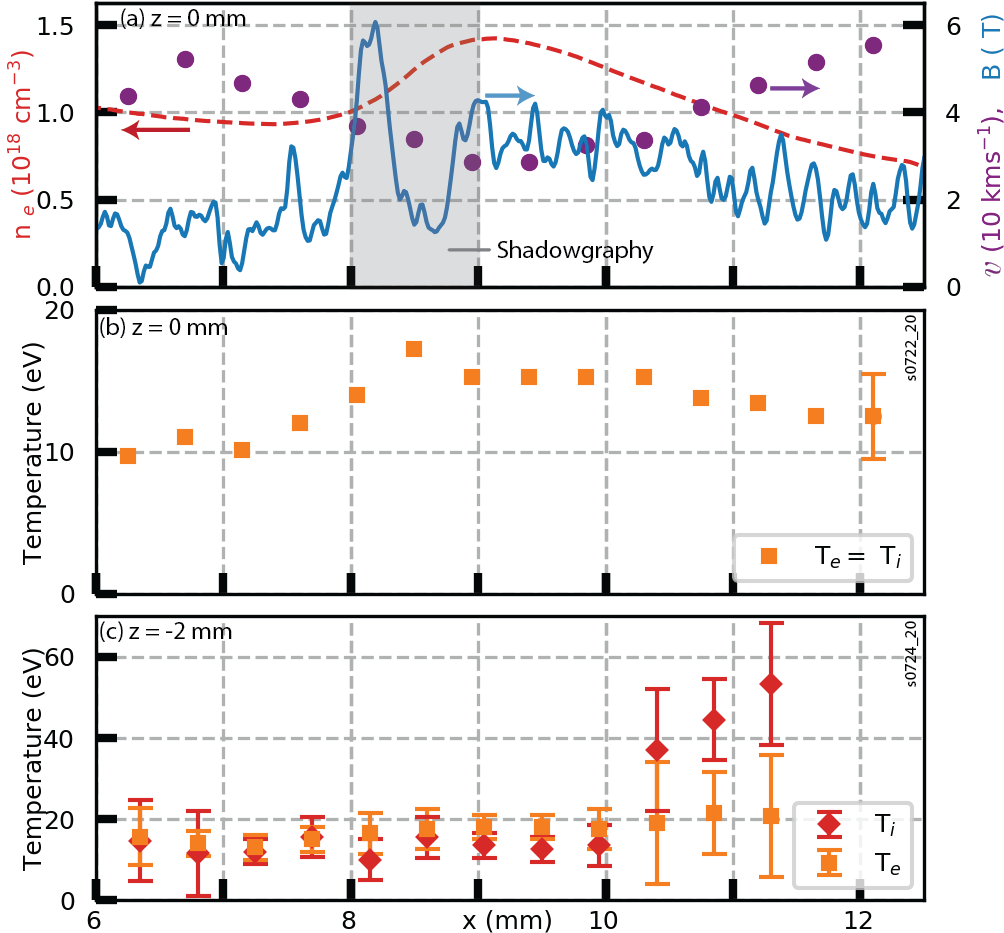} 
	\centering
	\caption{Thomson scattering data and plasma parameter profiles. (a) Flow velocity at $416$ ns with electron density and magnetic field lineouts at $z = 0$ mm. The region in which the magnetic field measurement was affected by shadowgraphy is shaded. (b) Temperature at $z = 0$ mm. A characteristic error bar which includes modeling uncertainty is shown for the final data point. (c) Ion and electron temperatures at $z = -2$ mm.}
	\label{fig:TS}
\end{figure}

Fig. \ref{fig:TS} (a) and (b) show the velocity and temperature measurements from TS in the $z = 0$ mm plane alongside lineouts of electron density and magnetic field (the locations of the TS scattering volumes are shown in Fig. \ref{fig:int_far} (a)). The electron density increases across the subcritical shock before decreasing as the plasma expands into the vacuum behind the shock. The velocity, which was measured locally, is consistent with 1D conservation of mass, suggesting that line integration does not affect the interferometry result. The ion and electron temperatures are equal across the shock and change by less than 10 eV ($T = T_{e} = T_{i}$ is presented in Fig. \ref{fig:TS} (b)).

We characterise the subcritical shock in terms of the density compression ratio and the shock width. The compression ratio, $R$, is estimated from electron density measurements. The compression ratio is estimated by comparing the downstream density in experiments with obstacles to the density at the same $x$ location in an experiment without obstacles (null shot), which yields $R = 2.7 \pm 0.5$. This is consistent with the MHD shock jump conditions for all reasonable values of $\gamma$.

The measured shock width, defined as the distance between $10\%$ and $90\%$ of the density jump, was $0.87 \pm 0.08$ mm. The width of a shock is determined by the dissipative and/or dispersive processes which increase entropy and transport energy at the shock front \cite{Kennel1984}. Table \ref{tab:Al_Cu} shows a comparison of the shock width with characteristic length scales for viscous dissipation ($\lambda_{i,i}$), Ohmic dissipation ($L_{\eta}$), electron heat conduction ($L_{\chi}$), and the formation of a cross shock potential due to two-fluid effects ($d_{i}$).

The ion-ion mean free path is $\sim 4$ orders of magnitude smaller than the shock width. This indicates that viscous dissipation does not shape the shock structure and that the shock is subcritical. Resistive diffusion, heat conduction and two-fluid effects may all contribute to the shaping of a subcritical shock and their characteristic scale lengths are all comparable to the shock width. The largest dissipative scale, and the scale closest to the shock width is $L_{\eta}$. This suggests that Ohmic dissipation plays the most significant role in shock shaping. Since $L_{\eta}$ is approximately equal to the shock width, the shock structure can be described by classical (Spitzer) resistive MHD only, without including anomalous resistivity. The contribution of electron heat conduction to shock shaping will be less than that of Ohmic dissipation since $L_{\eta} \sim 5 \times L_{\chi}$ in the upstream and $L_{\eta} > L_{\chi}$ across the entire subcritical shock. Two-fluid effects may also contribute to shock structure since $d_{i}$ is also approximately equal to the shock width. The formation of a cross shock potential due to two-fluid separation is a dispersive effect and does not dissipate kinetic energy. Rather, it excites whistler waves which carry energy away from the shock front. Since the dispersive scale, $d_{i}$, is approximately equal to the largest dissipative scale, $L_{\eta}$, this energy will be quickly dissipated and will not result in shock oscillations typical of collisionless systems. We note that while heat conduction and two-fluid effects may well contribute to shock structure, resistive diffusion alone is sufficient to explain the observed shock width.

\begin{table}
	\begin{tabular}{ p{5cm} p{1.3cm} p{1.8cm}}
	\hline
	\hline
	Parameter &  & Value (mm) \\ 
	\hline
	Shock width & & $0.87 \pm 0.08$  \\
	Ion-ion m.f.p & $\lambda_{i,i}$ & $8 \times 10^{-5}$  \\
	Resistive diffusion length & $L_{\eta}$ & $0.75$  \\
	Electron thermal diffusion length & $L_{\chi}$ & $0.16$  \\
	Ion inertial length & $d_{i}$ & $0.69$  \\
	\hline
	\hline
	\end{tabular}
	\centering
	\caption{Comparison of the measured shock width with characteristic dissipative and dispersive scale lengths.}
	\label{tab:Al_Cu}
\end{table}

\begin{figure}
	\includegraphics[scale=1]{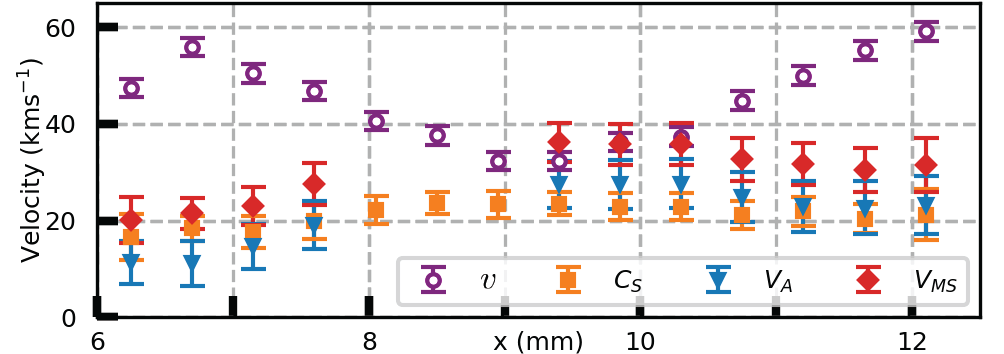} 
	\centering
	\caption{Flow velocity compared with the sound speed, $C_{S}$, Alfv\'{e}n speed, $V_{A}$, and fast-magnetosonic speed, $V_{MS}$, across the subcritical shock. Values which depend on $B$ are excluded for $x = 8 - 9$ mm where Faraday rotation measurements are affected by shadowgraphy.}
	\label{fig:velocity}
\end{figure}

The sound speed, Alfv\'{e}n speed and fast-magnetosonic speed are compared with the flow velocity in Fig. \ref{fig:velocity}. The flow becomes sub-magnetosonic across the shock but remains supersonic, a key signature of a subcritical shock. As the flow passes the obstacles, it expands into the vacuum and becomes super-magnetosonic again.

Since viscous dissipation does not shape the subcritical shock, the main heating mechanisms will be adiabatic and Ohmic. We estimate the heating due to adiabatic compression by calculating $T_{2} = T_{1} \times R^{\gamma -1} = 23 \pm 3$ eV for $\gamma = 5/3$. Ohmic heating will also increase the temperature and the heating power per unit volume can be estimated by

\begin{equation}
	P = \eta J^{2} = \eta \left( \frac{c}{4\pi} \right)^{2} | \nabla \times B |^{2},\\
	\label{heating_power}
\end{equation}

where $\eta$ is the Spitzer resistivity and $\nabla \times B \approx \Delta B_{y}/\Delta x \approx 2$ T$/1$ mm. This yields an increase in electron temperature of $\sim 20$ eV at the subcritical shock (assuming a velocity of $40$ \si{\kilo\metre\per\second}). However, we note that the radiative cooling time ($\sim 10$ ns \cite{Suzuki-Vidal2015a}) is less than the time the plasma takes to cross the shock ($\sim 25$ ns) so radiative cooling will reduce the observed temperature change. The estimated heating rate due to compression and Ohmic heating is $ \sim 40$ eV$/25$ ns $= 1.6$ eV\si{\per\nano\second}. Balancing this against the radiative cooling rates presented in Ref. \cite{Suzuki-Vidal2015a} requires an electron temperature of $\sim 15$ eV, in good agreement with the experimental results.

Behind the subcritical shock, the stagnation shock wings are stationary and remain downstream of the subcritical shock for the duration of the experiment. This suggests that the relevant Mach number for the stagnation shocks is the sonic Mach number, since this remains greater than unity across the subcritical shock. In this case, the stagnation shocks should be hydrodynamic jumps, formed by viscous dissipation on a spatial scale comparable to $\lambda_{i,i}$. This is below the resolution of the interferometry diagnostic ($\sim 0.5$ mm), but TS measurements provide strong evidence for this interpretation. Fig. \ref{fig:TS} (c) shows TS temperature measurements collected in the $x$-$y$ plane at $z = -2$ mm which cross a stagnation shock (see locations of scattering volumes in Fig. \ref{fig:int_far} (a)). The measurements are consistent with those at $z = 0$ mm in the upstream and subcritical shock, but show substantial ion heating to $40 - 70$ eV at the stagnation shock. The flow velocity decreases to $\sim 10$ \si{\kilo\metre\per\second} in this region and becomes subsonic. Both the ion heating and the subsonic downstream flow indicate that, in contrast to the subcritical shock, viscous dissipation of kinetic energy into ion thermal energy shapes the stagnation shocks.

To further investigate the role of resistive diffusion in shaping the subcritical shock, 2D simulations were carried out using AstroBEAR \cite{Cunningham2009,Carroll-Nellenback2013}, an adaptive mesh refinement (AMR) MHD code including resistivity and radiative loss. The two obstacles were simulated as $4$ mm diameter perfectly conducting internal reflection boundaries with a centre-to-centre separation of $9$ mm and were $20$ mm from the left hand wall. A constant plasma flow was injected from the this wall and the three remaining boundaries were zero gradient boundaries. Two levels of AMR gave an effective resolution of $0.125$ mm. The properties of the plasma wind, including resistivity, were varied to study the physics of the observed shocks. Radiative cooling was implemented by using the lookup table for aluminium presented in Ref. \cite{Suzuki-Vidal2015a}. We study the shocks which formed $500$ ns after the initial wind injection. By this time the shocks were well established and slow moving in the obstacle frame.

\begin{figure}
	\includegraphics[scale=1]{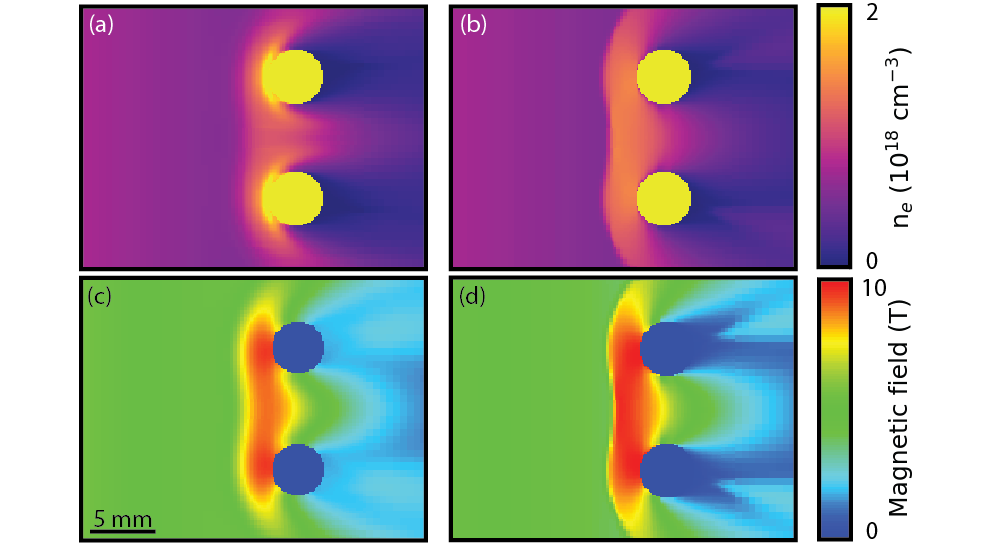} 
	\centering
	\caption{Simulation results at $500$ ns showing electron density and magnetic field with (a, c) $L_{\eta} = 0.5$ mm and (b, d) $L_{\eta} = 0$ mm.}
	\label{fig:sims}
\end{figure}

Fig. \ref{fig:sims} shows results from two different simulations. In (a, c) the resistive diffusion length was $0.5$ mm upstream of the shock. The upstream plasma parameters were $B = 4.5$ T, $n_{e} = 0.6 \times 10^{18}$ \si{\per\centi\metre\cubed}, $v = 70$ \si{\kilo\metre\per\second}, $T = 7$ eV and $M_{MS} = 1.5$ (comparable to the experimental value). The simulation reproduced the morphology of the experiment, with a smooth and continuous subcritical shock forming upstream of both obstacles. The shock width was $2.2$ mm. This suggests that, as in the experiments, viscous dissipation does not shape the shock. In contrast, (b, d) show a simulation with the same initial conditions but $\eta = 0$. In this case $\lambda_{i,i}$ exceeds $L_{\eta}$ (which is zero) so the shock width is set by viscous dissipation and is limited by the simulation resolution. This shows that resistive diffusion does indeed set the shock width in (a, c), with a shock width comparable to the classical resistive diffusion length.

In summary, we have presented an investigation of perpendicular subcritical shocks in a collisional plasma with $M_{MS} \sim 1.9$ and $\lambda_{i,i} \ll L_{\eta}$. We demonstrate that the shock is subcritical by showing that $M_{S} > 1$ across the shock. We confirm the theoretically predicted absence of a hydrodynamic jump, and show that the shock width is approximately equal to the classical (Spitzer) resistive diffusion length. We observe little heating at the subcritical shock, which is consistent with an absence of viscous dissipation. In contrast, downstream stagnation shocks cause substantial ion heating and produce a subsonic downstream flow. We interpret these shocks to be hydrodynamic in nature. Two-dimensional resistive MHD simulations reproduce the experimentally observed morphology of the subcritical shocks and demonstrate that Ohmic dissipation sets the shock width, which is comparable to the classical resistive diffusion length.

\begin{acknowledgments}
This work was supported by First Light Fusion Ltd. and by the US Department of Energy (DoE) including awards no. DE-NA0003764 and DE-SC0020434.
\end{acknowledgments}

\bibliography{Paper}

\end{document}